%
%

\magnification=\magstep2

\hsize=6.47truein
\vsize=8.89truein
\hoffset=-0.36truein
\voffset=0.01truein

\pageno 1

\baselineskip=15.6pt

\def\gsim{\,$\raise0.3ex\hbox{$>$}\llap{\lower0.8ex\hbox{$\sim$}}$\,}
\def\lsim{\,$\raise0.3ex\hbox{$<$}\llap{\lower0.8ex\hbox{$\sim$}}$\,}
\def\vecS{{\vec S}}

\vskip 10.0pt

\noindent
Version date: August 8, 1997 (final version)

\vskip 10.0pt

\noindent
\centerline{\bf Ground-state magnetization curve of}

\centerline{\bf a generalized spin-1/2 ladder}

\vskip 10.0pt

\noindent
\centerline{Takashi Tonegawa$^1$, Takeshi Nishida$^1$ and Makoto Kaburagi$^2$}

\vskip 10pt

\noindent
$^1${\it Department of Physics, Faculty of Science, Kobe University, Rokkodai,
Kobe 657, Japan}

\noindent
$^2${\it Department of Informatics, Faculty of Cross-Cultural Studies,
Kobe University, Tsurukabuto, Kobe 657, Japan}

\vskip 10.0pt

Employing a method of exact diagonalization for finite-size systems, we
investigate the magnetization curve in the ground state of an
antiferromagnetic spin-1/2 ladder with additional exchange interactions on
diagonal bonds, which is equivalent to an antiferromagnetic spin-1/2 chain
with bond-alternating nearest-neighbor and uniform next-nearest-neighbor
interactions.  It is found that a half-plateau appears in the magnetization
curve in a certain range of the interaction constants.  This result is
discussed in connection with the necessary condition for the appearance of the
plateau, recently given by Oshikawa et al.

\vskip 10.0pt

\parindent=4.7pc
\item{Key Words:} ground-state magnetization curve, generalized spin-1/2
ladder, spin-1/2 chain with competing interactions, half-plateau

\vskip 10.0pt

\noindent
Address for Further Correspondence:

\parindent=1.5pc
\item{} Takashi Tonegawa

\itemitem{} Department of Physics, Faculty of Science, Kobe University,

\itemitem{} Rokkodai, Kobe 657, JAPAN

\itemitem{} Tel.: +81-78-803-0541$\quad\quad$ Fax: +81-78-803-0722

\itemitem{} e-mail: tonegawa@kobe-u.ac.jp

\vfill\eject

\baselineskip=16.7pt

\parindent=1.5pc
There has been a considerable current interest in the study of quantum spin
systems with competing interactions, which exhibit a variety of fascinating
phenomena originating from frustration and quantum fluctuation.  In this
paper we investigate the magnetization curve in the ground state of an
antiferromagnetic spin-1/2 ladder with additional exchange interactions on
diagonal bonds.  We express the Hamiltonian describing this system in an
external magnetic field as
$$ \eqalignno{
  {\cal H} &= {\cal H}_0 + {\cal H}_{\rm Z}\,,     &(1{\rm a})            \cr
 &{\cal H}_0
  = (1 + \alpha) \sum_{\ell=1}^{N/2} \vecS_{1,\ell}\cdot\vecS_{2,\ell}
  + J \sum_{j=1}^2 \sum_{\ell=1}^{N/2} \vecS_{j,\ell}\cdot\vecS_{j,\ell+1}
                                                                          \cr
 &\qquad\qquad
   + (1 - \alpha) \sum_{\ell=1}^{N/2} \vecS_{1,\ell}\cdot\vecS_{2,\ell+1}\,,
                                                   &(1{\rm b})            \cr
 &{\cal H}_{\rm Z} = - H \sum_{j=1}^2 \sum_{\ell=1}^{N/2} S_{j,\ell}^z\,,
                                                   &(1{\rm c})            \cr}
$$
where $\vecS_{j,\ell}$ is the spin-1/2 operator at the $j$th leg and the
$\ell$th rung; \hbox{$1\!+\!\alpha$} and \hbox{$1\!-\!\alpha$}
(\hbox{$0.0\!\leq\!\alpha\!\leq\!1.0$}) are, respectively, the interaction
constants between neighboring spins along the rung and diagonal bonds; $J$
($J\!\geq\!0.0$) is the interaction constant between neighboring spins along
the leg bond; $H$ ($H\!\geq\!0.0$) is the magnitude, in an appropriate unit,
of the magnetic field applied along the $z$-axis; $N$ is the total number of
spins in the system and is assumed to be a multiple of four.  We impose
periodic boundary conditions ($\vecS_{j,N+1}\!\equiv\!\vecS_{j,1}$).  It
should be noted that this system is equivalent to an antiferromagnetic
spin-1/2 chain with both bond-alternating nearest-neighbor interactions, the
interaction constants being \hbox{$1\!+\!\alpha$} and \hbox{$1\!-\!\alpha$},
and uniform next-nearest-neighbor interactions, the interaction constant being
$J$, which compete with each other.

\parindent=1.5pc
The ground-state [1] and thermodynamic [2-4] properties of the system in the
case of \hbox{$\alpha\!=\!0.0$} and \hbox{$H\!=\!0.0$} have been studied
extensively.  In particular, it has been found that as the value of $J$
increases, the phase transition from the massless spin-fluid phase to the
massive dimer phase occurs at $J\!=\!J_{\rm c}\!\sim\!0.2412$ in the ground
state [4-6].  As is well known, the exact ground-state magnetization curve in
the case of $\alpha\!=\!J\!=\!0.0$ has been obtained by Griffiths [7].  The
ground-state magnetization curves in the case of $\alpha\!=\!0.0$ [8] and in
the case of $J\!=\!0.0$ [9] have also been numerically calculated for various
values of $J$ and $\alpha$, respectively.

\parindent=1.5pc
Obtaining the ground-state magnetization curve of the present system, we
employ Sakai and Takahashi's method [10], by the use of which they have
discussed the ground-state magnetization curve for a uniform spin-1
chain.  The outline of this method, which is based on a method of exact
diagonalization for finite-size systems combined with the conformal field
theory, can be summarized as follows.  Let $E_0(N,M)$ be the lowest-energy
eigenvalue, within the subspace determined by the
value $M\!=\!\sum_{j=1}^{2}\sum_{\ell=1}^{N/2}S_{j,\ell}^{z}(=\!0$, $1$,
$\cdots$, $N/2$), of the Hamiltonian ${\cal H}_0$ for a given $N$.  Then, the
conformal field theory predicts that, if the state with the energy $E_0(N,M)$
is massless, the asymptotic behavior of $E_0(N,M)$ in the thermodynamic
(\hbox{$N\!\to\!\infty$}) limit has the form [11]
$$
   {E_0(N,M) \over N}
      \sim \varepsilon(m) - C(m) {1 \over N^2}\quad\quad(N\to\infty)
                                                                   \eqno (2)
$$
with \hbox{$m\!\equiv\!M/N$}, where $\varepsilon(m)$ is the lowest energy per
spin for a given $m$ in the thermodynamic limit and $C(m)$ is a positive
constant which is proportional to the product of the central charge and the
sound velocity.  Minimizing with respect to $m$ the total energy,
\hbox{$\varepsilon_{\rm tot}\!=\!\varepsilon(m)\!-\!Hm$}, per spin of the
system described by the Hamiltonian ${\cal H}$, we can obtain the equation
which relates the ground-state magnetization $\langle m \rangle$ per
spin in the thermodynamic limit with $H$ as
$$
     \varepsilon'\bigl(\langle m \rangle\bigr) = H\;.               \eqno (3)
$$
From Eq.$\,$(2) we can derive
$$ \eqalignno{
 \Delta E_0(N;M,M-1)
  & \sim \varepsilon'(m_{-0})
        - {1\over 2}\varepsilon''(m_{-0}) {1\over N}
                \quad\quad(N\to\infty)\;,              &(4{\rm a})    \cr
 \Delta E_0(N;M+1,M)
  & \sim \varepsilon'(m_{+0})
        + {1\over 2}\varepsilon''(m_{+0}) {1\over N}
                \quad\quad(N\to\infty)\;,              &(4{\rm b})    \cr}
$$
where
\hbox{$\Delta E_0(N;M,M\!-\!1)\!\equiv\!E_0(N,M)\!-\!E_0(N,M\!-\!1)$}.  The
essential point of Sakai and Takahashi's method [10] is that as
far as the massless states are concerned, Eqs.$\,$(4a,b) hold when $N$ is
sufficiently large, and thus the value \hbox{$H\!=\!\varepsilon'(m)$}
[see Eq.$\,$(3)] of the magnetic field for a given $m$
(\hbox{$0\!<\!m\!<\!1/2$}) can be estimated by making an extrapolation
which uses these equations to be
\hbox{$H\!=\!\varepsilon'(m_{-0})\!=\!\varepsilon'(m_{+0})$}.  For the
estimation of \hbox{$H_{{\rm c}0}\!\equiv\!\varepsilon'(0)$}, which is the
critical field at which $\langle m \rangle$ starts to increase from zero in
the ground-state magnetization curve, we apply Shanks' transformation [12] to
the sequences \hbox{$\{\Delta E_0(N;1,0)\}$}, following Sakai and
Takahashi's procedure [13].  Furthermore, the saturation
field $H_{{\rm s}}\!\equiv\!\varepsilon'(1/2)$ can be obtained analytically,
since it is straightforward to diagonalize the Hamiltonian ${\cal H}_0$
within the \hbox{$M\!=\!(N/2)\!-\!1$} subspace; it is given by
\hbox{$H_{{\rm s}}\!=\!{\rm Max}(2,\,1\!+\!2J\!+\alpha)$} in the present
case.  Recently, we have successfully applied the method to the case of an
antiferromagnetic spin-1 chain with bond-alternating nearest-neighbor
interactions and uniaxial single-ion-type anisotropy [14].

\parindent=1.5pc
We obtain the ground-state magnetization curve in the thermodynamic limit for
$J\!=\!0.1$, $0.2$, $0.3$, and $0.4$, for each of which various values of
$\alpha$ are chosen.  In the calculation we numerically diagonalize the
Hamiltonian ${\cal H}_0$, using the computer program package KOBEPACK/S~[15],
to calculate $E_0(N,M)$ for \hbox{$N\!=\!8$}, $12$, $\cdots$, $24$.  Then, we
can make the analysis for $m\!=\!1/12$, $1/8$, $1/6$, $1/4$, $1/3$, $3/8$, and
$5/12$.  Our calculation shows that both $\Delta E_0(N;M\!+\!1,M)$ and
$\Delta E_0(N;M,M\!-\!1)$ are almost linear functions of $1/N$ at least for
$m\!=\!1/8$ and $3/8$ in accordance with the forms given by
Eqs.$\,$(4a,b).  The values of $\varepsilon'(m_{-0})$ and
$\varepsilon'(m_{+0})$ for these $m$'s can thus be estimated.  In a
similar way we estimate $\varepsilon'(m_{-0})$ and $\varepsilon'(m_{+0})$ for
$m\!=\!1/12$, $1/6$, $1/3$, and $5/12$, assuming Eqs.$\,$(4a,b), although only
two data are available.  All the obtained results show that
$\varepsilon'(m_{-0})$ and $\varepsilon'(m_{+0})$ coincide with each other
within the numerical error.  For $m\!=\!1/4$, on the other hand,
Eqs.$\,$(4a,$\,$b) do or do not hold depending upon the values of $\alpha$
and $J$.  The case where Eqs.$\,$(4a,$\,$b) do not hold is the case where the
state with $m\!=\!1/4$ is massive and therefore $\varepsilon'(1/4_{-0})$ is
smaller than $\varepsilon'(1/4_{+0})$.  This means that, in the magnetization
curve, there appears the
\hbox{{\rm half}$\bigl(\langle m \rangle\!=\!1/4\,\bigr)$-plateau}
with the critical field \hbox{$H_{{\rm c}1}\!\equiv\!\varepsilon'(1/4_{-0})$}
at which the plateau starts and that
$H_{{\rm c}2}\!\equiv\!\varepsilon'(1/4_{+0})$ at which it ends.  We estimate
the former and latter critical fields by applying Shanks' transformation [12]
to the sequences \hbox{$\{\Delta E_0(N;N/4,N/4\!-\!1)\}$} and
\hbox{$\{\Delta E_0(N;N/4\!+\!1,N/4)\}$}, respectively.

\parindent=1.5pc
We find that the half-plateau appears in the magnetization curve when
\hbox{$0.5\lsim\alpha\lsim0.95$} for $J\!=\!0.1$, when
\hbox{$0.2\lsim\alpha\lsim0.85$} for $J\!=\!0.2$, when
\hbox{$0.0\lsim\alpha\lsim0.8$} for $J\!=\!0.3$, and when
\hbox{$0.0\!<\!\alpha\lsim0.75$} for $J\!=\!0.4$.  As an example, we depict in
Fig.$\,$1 the magnetization curve with the
half-plateau, obtained for $J\!=\!0.2$ and
$\alpha\!=\!0.5$.  Plotting versus $\alpha$ the critical fields $H_{{\rm c}0}$,
$H_{{\rm c}1}$, and $H_{{\rm c}2}$ as well as the saturation field
$H_{{\rm s}}$, we can draw the ground-state phase diagram on the $H$ versus
$\alpha$ plane; the result for $J\!=\!0.2$ is shown in Fig.$\,$2.

\parindent=1.5pc
We also calculate numerically the eigenfunctions of the lowest- and
second-lowest-energy states within the \hbox{$M\!=\!N/4$} subspace for the
finite-$N$ systems, and find that at least for a set of $J$ and $\alpha$
giving the  plateau, the lowest-energy state for \hbox{$M\!=\!N/4$} in the
thermodynamic limit is doubly degenerate, one of the eigenfunctions of which
has the periodicity $n\!=\!4$ (in units of the lattice constant) concerning
the translational symmetry.  This result is consistent with the necessary
condition \hbox{$n\bigl(S\!-\!\langle m \rangle\bigr)\!={\rm integer}$} for
the appearance of the plateau with the magnetization $\langle m \rangle$ ($S$
is the magnitude of spins), which has recently been given by Oshikawa,
Yamanaka, and Affleck [16].  Finally, it should be noted that very recently
Totsuka [17] has clarified in an excellent way the mechanism for the
appearance of the plateau in the present system, using a bosonization
technique.  According to this work, it becomes clear that the
next-nearest-neighbor interaction plays a crucial role in the appearance of
the plateau.

\vskip 16.7pt

\noindent
{\bf Acknowledgements}

We would like to thank Drs.~M.~Hagiwara, K.~Totsuka, and M.~Yamanaka for
invaluable discussions.  We also thank the Supercomputer Center, Institute for
Solid State Physics, University of Tokyo and the Computer Center, Tohoku
University for computational facilities.  The present work has been supported
in part by a Grant-in-Aid for Scientific Research from the Ministry of
Education, Science, Sports and Culture, Japan.

\vfill\eject

\noindent
{\bf References}

\parindent=2.0pc
\item{[1]} For a review, see I.~Harada and T.~Tonegawa,~{\it Recent Advances
in Magnetism of Transition Metal Compounds}, ed.~A.~Kotani and N. Suzuki
(World Scientific, Singapore, 1993) p.$\,$348.

\item{[2]} I.~Harada, T.~Kimura and T.~Tonegawa,~J.~Phys.~Soc.~Jpn.~{\bf 57}
(1988) 2779; T.~Tonegawa and I.~Harada,~J.~Phys.~Soc.~Jpn.~{\bf 58} (1989)
2902.

\item{[3]} J.~Riera and A.~Dobry,~Phys.~Rev.~B {\bf 51} (1995) 16098.

\item{[4]} G.~Castilla, S.~Chakravarty and
V.~J.~Emery,~Phys.~Rev.~Lett.~{\bf 75} (1995) 1823.

\item{[5]} K.~Okamoto and K.~Nomura,~Phys.~Lett.~A {\bf 169} (1992) 433.

\item{[6]} S.~Eggert,~Phys.~Rev.~B {\bf 54} (1996) R9612.

\item{[7]} R.~B.~Griffiths,~Phys.~Rev.~{\bf 133} (1964) A768.

\item{[8]} T.~Tonegawa and I.~Harada,~J.~Phys.~Soc.~Jpn.~{\bf 56} (1987)
2153; see also T.~Tonegawa and I.~Harada, Physica B {\bf 155} (1989) 379.

\item{[9]} J.~C.~Bonner, S.~A.~Friedberg, H.~Kobayashi, D.~L.~Meier and
H.~W.~J. Bl{\"o}te,~Phys.~Rev.~B {\bf 27} (1983) 248.

\item{[10]} T.~Sakai and M.~Takahashi,~Phys.~Rev.~B {\bf 43} (1991) 13383.

\item{[11]} J.~L.~Cardy,~J.~Phys.~A {\bf 17} (1984) L385; H.~W.~J.~Bl{\"o}te,
J.~L.~Cardy and M.~P.~Nightingale,~Phys.~Rev.~Lett.~{\bf 56} (1986) 742;
I.~Affleck, Phys.~Rev.~Lett.~{\bf 56} (1986) 746.

\item{[12]} D.~Shanks,~J.~Math.~Phys.~{\bf 34} (1955) 1.

\item{[13]} T.~Sakai and M.~Takahashi,~Phys.~Rev.~B {\bf 42} (1990) 1090.

\item{[14]} T.~Tonegawa, T.~Nakao and M.~Kaburagi,~J.~Phys.~Soc.~Jpn.~{\bf 65}
(1996) 3317.

\item{[15]} M.~Kaburagi, T.~Tonegawa and T.~Nishino, {\it Computational
Approach- es in Condensed Matter Physics}, Springer Proc.~Phys.~{\bf 70},
ed.~S.~Miya- shita, M.~Imada and H.~Takayama (Springer-Verlag, Berlin, 1992)
p.$\,$179.

\item{[16]} M.~Oshikawa, M.~Yamanaka and I.~Affleck,~Phys.~Rev.~Lett.~{\bf 78}
(1997) 1984.

\item{[17]} K.~Totsuka,~preprint.

\vfill\eject

\noindent
{\bf Figure Captions}

\vskip 10.0pt

\noindent
Fig.$\,$1.~~Ground-state magnetization curve in the thermodynamic limit
obtained for $J\!=\!0.2$ and $\alpha\!=\!0.5$.  The closed circles show plots
of $\langle m \rangle$ versus $H\!=\!\varepsilon'(\langle m \rangle)$.  The
solid lines are guides to the eye.

\vskip 10.0pt

\noindent
Fig.$\,$2~~Ground-state phase diagram on the $H$ versus $\alpha$ plane in the
thermodynamic limit for \hbox{$J\!=\!0.2$}.   The closed circles show plots
versus $\alpha$ of the critical fields $H_{{\rm c}0}$, $H_{{\rm c}1}$, and
$H_{{\rm c}2}$ and also of the saturation field $H_{\rm s}$.  The solid lines
are guides to the eye.  The magnetization $\langle m \rangle$ is given by
\hbox{$\langle m \rangle\!=\!0$}, $1/4$, and $1/2$ in the regions A, B, and
C, respectively.  In the remaining region $\langle m \rangle$ increases
continuously as $H$ increases.

\bye